\input{epsf}
\documentstyle[aps,preprint,epsf]{revtex}

\tightenlines
\begin{document}

\title{The Conformal Mode in 2D Simplicial Gravity}
\vspace{1.5 true cm}
\author{Simon Catterall}
\address{Physics Dept., Syracuse Univ.,\\
Syracuse, NY 13244\\
smc@suhep.phy.syr.edu}
\author{Emil Mottola}
\address{Theoretical Division T-8,\\
Los Alamos National Laboratory,\\
Los Alamos, NM 87545\\
emil@lanl.gov}

\date{\today}
\preprint{SU-4240-700}
\maketitle

\begin{abstract}
\baselineskip 1 pc
We verify that summing 2D DT geometries correctly reproduces the Polyakov
action for the conformal mode, including all ghost contributions, at
large volumes. The Gaussian action is reproduced even for $c_m=10$,
well into the branched polymer phase, which confirms the expectation that the
DT measure is indeed correct in this regime as well
\thispagestyle{empty}
\end{abstract}
\voffset=1.0 true cm
\newpage 

\voffset=-0.5 true cm
\pagestyle{plain}
\pagenumbering{arabic}

\newcommand{\sq}{\lower.15ex\hbox{\large$\Box$}}
\newcommand{\sqb}{\lower.25ex\hbox{\large$\stackrel
{\mbox{\rule[.6mm]{2.4mm}{.13mm}}}{\Box}$}}

\noindent {\it Introduction.}
In the last decade there has been considerable progress in our
understanding of two dimensional quantum gravity (2DQG). The key 
element that has made this progress possible is the recognition that
the trace anomaly requires the effective action of 2DQG 
to be augmented by a well-defined nonlocal term \cite{Poly},
\begin{equation}
S_{anom} = {Q^2 \over 16 \pi}\int d^2x \sqrt {-g}\int d^2x'\sqrt{-g'}
\ R(x)\ \sq ^{-1}(x,x') \ R(x')\ .
\label{Sanom}
\end{equation}
Because this action is conformally invariant, and the massless scalar
propagator $\sq ^{-1}(x,x') \sim \log (x-x')^2$ in two dimensions, $S_{anom}$
is relevant (strictly, marginal) in both the infrared and ultraviolet, and
leads to nontrivial scaling of the distribution of random geometries in the
path integral for 2DQG. 

This scaling behavior has been found analytically in the 
continuum, both by canonical operator methods for
the current algebra of the energy-momentum tensor \cite{KPZ}, and
by covariant conformal field theory methods for the Euclidean
correlation functions \cite{DDK}. The former KPZ approach exposes the
relation of the parameter $Q^2$ to the unique central extension 
of the current algebra of diffeomorphisms on the 2D world sheet,
\begin{equation}
Q^2 = {25 - c_m\over 6}\ ,
\label{cent}
\end{equation}
where $c_m$ is the central charge of the matter representation.
For free matter $c_m$ is the integer number of free boson
or fermion fields: $c_m = N_s + N_f$. The value of $Q^2 = {25\over 6}$ 
for $c_m =0$ is the value for pure 2DQG.

The second covariant approach to 2DQG is more 
closely related to the statistical properties of the ensemble of 2D
geometries, and may be checked by a numerical Monte Carlo method applied 
to a discretized Euclidean worldsheet \cite{DT}. In this approach fluctuations
of the worldsheet are modelled by performing a sum over a set of dynamical
triangulations (DTs). The DT numerical simulations
agree with the continuum scaling predictions in all important respects
\cite{dtrev}. In
particular, correlation functions of conformal operators acquire well-defined
anomalous scaling dimensions, which can be computed in terms of $Q^2$ and
agree with (\ref{cent}), including the miminal models for which
$c_m = 1- 6/n(n+1) <1$, where $n$ is an integer $n \ge 2$. What is remarkable
about this agreement is that the anomalous action (\ref{Sanom}) which gives
rise to the `gravitational dressing' of conformal operators in the continuum
is not inserted into the discretized action of the DT simulation by hand, but
instead must be generated {\it automatically} with the correct coefficient by
the integration measure on the space of geometries in the DT approach.
 
Despite the satisfactory agreement between theory and the DT approach for
correlation functions in 2DQG, one would like to demonstrate
explicitly the appearance of $e^{S_{anom}}$ in the measure of 2DQG,
particularly for $c_m > 1$, where the continuum scaling relations become
complex and the interpretation of the theory becomes much less clear. 
This is our main purpose in this Letter.
  
For studying this question in more detail it is
useful to introduce the conformal parameterization of the metric,
\begin{equation}
g_{ab} = e^{2 \sigma} \bar g_{ab}\ ,
\label{conf}
\end{equation}
where $\bar g_{ab}$ is a fixed fiducial metric which depends only
on the global topology of the two dimensional Euclidean manifold.
Since all two-manifolds are locally conformally flat, there is
a local coordinate patch around every point where the fiducial
metric has the form $\delta_{ab}$. A finite number of such 
coordinate patches and global (Teichmuller) moduli parameters are
required to describe the fiducial metric in any given topology,
but all local deformations of $g_{ab}$ are contained in the conformal
factor $e^{2\sigma (x)}$. The Ricci scalar is given by 
\begin{equation}
R = e^{-2\sigma}(\overline R - 2\sqb \sigma) = e^{-2\sigma}\overline R 
- 2\sq \sigma
\label{Ric}
\end{equation}
in the parameterization (\ref{conf}), where the overbars refer
to the fiducial metric $\bar g_{ab}$ and the absence of an overbar
refers to the full metric $g_{ab}$. It is a theorem in 2D
Riemannian geometry that for any smooth manifold with given continuous
$R(x)$, a unique solution to the nonlinear equation for $\sigma$ (\ref{Ric})
exists such that $\overline R$ is everywhere a specified constant \cite{Yam}.
This constant is necessarily positive for closed manifolds with no handles, 
{\it i.e.} with the topology of the sphere $S^2$, zero for closed manifolds
with one handle, {\it i.e.} with the topology of the torus $T^2$ and negative
for closed manifolds with more than one handle. The uniformization (Yamabe)
theorem assures us that we can construct the conformal factor 
$e^{\sigma (x)}$ uniquely for every 2D closed manifold of fixed topology. 

In the conformal parameterization and Euclidean signature
the action $S_{anom}$ becomes 
\begin{equation}
S_E[\sigma] = - S_{anom} = {Q^2\over 4 \pi} \int d^2 x \ 
\sqrt {{\bar g}}\ \bigl[\sigma (-\,\sqb )\,\sigma + \overline R \sigma \bigr]
+ const.
\label{loc}
\end{equation}
which is local and quadratic in $\sigma$. Hence in this variable the
distribution of geometries in the path integral for 2DQG is
the same as that of a free scalar field with a well-defined
normalization. As long as $Q^2$ is positive this Gaussian distribution
is bounded on closed Riemannian manifolds, as is clear
upon integration of the quadratic term in (\ref{loc}) by parts.

Smooth geometries and the classical limit to 2DQG are recovered
only in the large $Q^2$ limit, $Q^2 \to \infty$ ($c_m \to -\infty$),
where the fluctuations in the conformal factor become suppressed.
The $25$ in (\ref{cent}) for pure 2DQG can be understood in
the conformal gauge as arising from $26$ from the ghosts of Faddeev-Popov
gauge fixing and $-1$ for $\sigma$ field itself, which contributes to the 
anomalous action as would one additional matter scalar degree of freedom,
lowering the effective value of $Q^2$.
Although the KPZ scaling relations become complex for $c_m >1$,
nothing singular occurs in the action $S_E$ as 
$c_m$ is raised above $1$ and $Q^2$ is lowered below $4$,   
However, a heuristic argument based on the competition of action and
entropy of singular `spike' solutions to the classical equations following
from $S_E$ suggests that at the critical value $Q_{cr}^2 = 4$ the theory
undergoes a BKT-like phase transition \cite {BKT} to a phase dominated by
elongated extrusions of the 2D world sheet, since the entropy of these
configurations first exceeds their action at this value of $Q^2$ \cite{Cates}.
Hence for $Q^2 \le Q_{cr}^2$, the typical 2D geometry would be expected 
to be very far from smooth, resembling instead a multiply branched polymer.
This branched polymer phase is seen also in the DT simulations for $c_m > 1$
\cite{c.ge.1}.

The significance of the critical value, $c_m=1$ may also be understood from a
canonical perspective on Lorentzian signature 2D manifolds. In the absence of
matter the quadratic theory specified by the anomalous action
$S_{anom}$ is overconstrained. This is clear from the fact that
the metric is a $2\times2$ symmetric matrix which has three
independent components, $g_{00}, g_{11}$ and $g_{01}= g_{10}$. Two of 
these are isomorphic to pure gauge diffeomorphisms of the 2D worldsheet
and can be removed. This
 leaves one independent componen which can be
taken to be the
 conformal mode $\sigma$, but its dynamics is constrained by
the
 two first class constraints in the $(0 0)$ and $(0 1)$ components of the
Euler-Lagrange equations following from $S_{anom}$. These are the 2D
lapse and shift constraints to be imposed on the phase space
in the canonical treatment of a theory with diffeomorphism invariance. 
Hence there are finally $3-2-2 = -1$ local degrees of freedom in pure 
2DQG described by the Polyakov action (\ref{Sanom}), or in other words the
theory is overconstrained and possesses no local degrees of freedom. When free
matter fields are added the theory has $c_m -1$ local degrees of freedom and
the conformal mode first can fluctuate locally when $c_m \ge 1$, which
coincides with the BKT phase transition critical value, $Q^2 \le Q^2_{cr} =4$ 
In this branched polymer phase where the conformal spike configurations become
`liberated,' 2DQG becomes sensitive to its UV cutoff and a smooth continuum
limit at large distance scales presumably does not exist, unless higher
derivative UV relevant operators or an explicit UV cutoff are introduced.

In the DT discretized approach to 2DQG the UV cutoff is supplied by the 
lattice scale $a$, while the IR behavior is controlled by the total 
2D volume, {\it i.e.} the sum of areas of all the triangles in the 
ensemble. If the interpretation of the $c_m =1$ `barrier' given above 
is correct we would expect the distribution of $\sigma$ fields on the 
lattice to remain Gaussian with a well-defined width determined by 
$S_E$ in the large volume limit, even in the branched polymer phase where the
typical geometries are irregular on the lattice scale. 
To check this hypothesis requires finding the conformal factor by solving 
the Yamabe equation (\ref{Ric}) on each geometry in the DT ensemble and
reconstructing the distribution of $\sigma$ in the ensemble. Besides casting
some light on the case $c_m \ge 1$ the construction of this Gaussian
distribution in $\sigma$ would provide explicit confirmation
of the identity of the continuum and lattice approaches,
independently of any correlation functions or expectation values of
observables. Preliminary studies of this distribution were reported in Refs. 
\cite{AghMig,lat98}.

We have an additional motivation for the construction of this
distribution in $\sigma$ from the analysis of the conformal anomaly generated
action in 4D. At the infrared fixed  point of this 4D action the
conformal factor distribution is again predicted to be Gaussian with a
width determined by the analogous anomaly coefficient in 4D \cite{AntMot}.
In the conformal parameterization (\ref{conf}) the 4D action analogous to
(\ref{loc}) is
\begin{equation}
S_{E}[\sigma]_{D=4} = {Q^2 \over (4 \pi)^2}\int d^4 x \sqrt{g}\, \bigl[
\sigma  {\Delta}_4 \sigma + \hbox{$1\over 2$}\bigl(G - \hbox{$2 \over 3$} 
\sq R \bigr) \sigma\bigr]\ ,
\label{act} 
\end{equation}
with the contribution to the coefficient from the number of massless
conformal scalar ($N_S$), Weyl fermion ($N_{WF}$) and vector ($N_V$) fields
given by
\begin{equation}
Q^2_{D=4} = {1 \over 180}(N_S + \hbox{$11\over 2$} N_{WF} + 62 N_V - 28) +
Q^2_{grav}\ ,
\label{centf}
\end{equation}
and $Q^2_{grav}$ the contribution of gravitons. 
In (\ref{act}) $G$ is the Gauss-Bonnet invariant and
$\Delta_4 = \sq ^2 + 2R^{ab} \nabla_a \nabla_b - {\textstyle \frac{2}{3}} R 
{\sq}^2 + {\textstyle \frac{1}{3}} (\nabla^a R) \nabla_a$
is the unique fourth order conformally covariant operator in 4D which has
a logarithmic propagator, analogous to $\sq$ in 2D. The action (\ref{act}) 
is therefore relevant (strictly, marginal) in the IR and leads to nontrivial
scaling of the cosmological and Newtonian `constants' at large distance scales
\cite{fractal}. These scaling relations become complex for $Q^2_{D=4} < 8$,
analogous to $Q^2 < 4$ in 2D, with the important difference that additional
matter fields take us into the smooth phase in 4D but the irregular
branched polymer phase in 2D, since matter contributes positively in 
(\ref{centf}) but negatively in (\ref{cent}). Moreover, singular 
`spike' solutions also exist in this 4D conformal theory and their 
entropy exceeds their action at precisely the same critical value at 
which the scaling relations become complex \cite{spikes}. 

Since the DT program in 4D has not met with the
 same level of success
as in 2D, it would be an important check on the model to construct
the distribution of the conformal factors generated by the DT
algorithm to determine if it is consistent with the continuum trace anomaly
and action (\ref{act}) in 4D. A positive result would provide a method of
determining the contribution of the gravitons $Q^2_{grav}$ to the Gaussian
width as well. Clearly the situation is quite a bit more complicated in 4D
and will require a careful disentangling of the graviton degrees
of freedom in addition to $\sigma$, but the basic idea of the reconstruction of
the metric of each member of the DT ensemble is similar in higher dimensions,
and the method should be tested first in the 2D case which is the most
complete and best understood model of QG which we have at the present time.\\

\noindent {\it Dynamical Triangulations.}
In the dynamical triangulation approach to 2DQG one constructs
an ensemble of discretized geometries, each one of which consists
of regular two-simplices, {\it i.e.} equilateral triangles,
glued together at their edges.
If the length of any side of one of the triangles is $a$ 
then its area is $A_{\Delta} = \sqrt 3 a^2/4$. If $N_0 = N$ is the 
total number of vertices in the lattice then the total number of links 
is $N_1 = 3(N - \chi_{_E})$ and the total number of triangles is 
$N_2 = 2(N - \chi_{_E})$, where  
\begin{equation}
\chi_{_E} = {1\over 4\pi}\int\ d^2x\sqrt g \ R \to 
\sum_i \left(1 - {q_i\over 6}\right) = N_0 - {N_1\over 3}
\label{Euler}
\end{equation}
is the Euler character of the triangulation and $q_i$ is the
coordination number of the vertex $i$, {\it i.e.} the
number of neigboring vertices to which it is connected by a single link.
Hence $\sum_i q_i = 2N_1$ since each link is counted twice
in this sum. In the following we restrict ourselves to spherical 
topology $S^2$ for which $\chi_{_E} = 2$.

In the dual lattice the area of the cell associated with the
vertex $i$ is 
\begin{equation}
(d^2x \sqrt g)_i \to \tilde A_i = {q_i\over 3} A_{\Delta}\ ,
\label{area}
\end{equation}
and the scalar curvature concentrated at the vertex $i$ is
\begin{equation}
R_i \to {2\pi\over q_i A_{\Delta}}(6-q_i)\ ,
\label{Regge}
\end{equation}
which together are consistent with (\ref{Euler}). 
When $q_i=6$ the vertex has no curvature associated
with it, since $6$ equilateral triangles joined at the vertex
tile a local region of flat space.

The scalar Laplacian on the lattice site $i$ can be represented as a sum 
over its $q_i$ nearest neighbors $j(i)$ as \cite{ItzDro}
\begin{equation}
(-\sq \sigma)_i \to {2\sqrt 3 \over q_i A_{\Delta}} \sum_{j(i)} (\sigma_i -
\sigma_j) = {2\sqrt 3 \over q_i A_{\Delta}} \sum_j (q_i \delta_{ij} -
C_{ij})\sigma_j\ , 
\label{lap}
\end{equation}
where $C_{ij}$ is the coordination or adjacency matrix, equal to one 
if $i$ and $j$ are connected by a single link and zero otherwise. Hence the
Yamabe equation (\ref{Ric}) on the lattice becomes
\begin{equation}
2M_{ij}\sigma_j=\frac{2\pi}{3}\left(6-q_i\right) - \frac{\overline{R} q_i}{3}\,
A_{\Delta}\, e^{-2\sigma_i}
\label{disr}
\end{equation}
after multiplication through by $\tilde A_i$. Here, we have defined
the matrix,
\begin{equation}
M_{ij}=\frac{2}{\sqrt{3}}\left(q_i\delta_{ij}-C_{ij}\right)\ 
\label{mat}
\end{equation}
The left side of Eq. (\ref{disr}) vanishes when summed over $i$.
Hence we obtain the constraint,
\begin{equation}
8\pi = \frac{\overline R\, A_{\Delta}}{3} \sum_i q_i
e^{-2\sigma_i} = \overline R\, A_{\Delta}\, N_2\ ,
\label{cons}
\end{equation}
where the last relation follows from the lattice version of
$\int \sqrt g e^{-2\sigma} = \int \sqrt{\bar g} \to A_{\Delta} N_2$.
Thus if $\overline R$ is held constant in the continuum limit,
$N_2 \to \infty$ and $A_{\Delta} \sim 1/N_2 \to 0$. This implies that
we scale the lattice spacing to zero like the inverse square root
of the number of triangles. 

The ensemble of configurations for the fixed area partition function 
is generated by a Monte Carlo simulation which updates the geometry
at fixed large $N_2$ using link flips with unit probability, {\it i.e.}
no gravitational action is used to weight the configurations. To couple
matter we just dress the vertices with
$c_m$ free scalar fields coupled by the usual Gaussian action using
the scalar Laplacian (\ref{lap}). A combined 
heat bath and overrelaxation algorithm was used to simulate the scalars.
For each of the runs we generated $10,000$ configurations separated by 100
Monte Carlo sweeps. Lattices of size with $N_2 = 100,\, 200,\, 400,\, 800,\,$
and $1600$ triangles were used.
 
For each configuration in the ensemble the conformal mode was
constructed by solving (\ref{mat}) by a modified Newton-Raphson
iteration scheme. In order to compare the results with the
anomalous action $S_{anom}$ the conformal mode is decomposed
into eigenmodes of the round sphere Laplacian,
$-\sqb = -\frac{\sqrt{g}}{\sqrt{\overline{g}}}\sq= -e^{2\sigma} \sq $.
A symmetrized version of this operator on the lattice can be defined by 
\begin{equation} 
L_{ij} = {3\over A_{\Delta}} {e^{\sigma_i + \sigma_j}\over \sqrt{q_iq_j}}M_{ij} 
\end{equation} 
The spectrum $\{\lambda^{\ell}\}$ and eigenmodes $\{u^{\ell}_i\}$ of this
operator were computed, 
\begin{equation}
L_{ij} u^{\ell}_j = \lambda_{\ell}\, u^{\ell}_i\ ,
\end{equation}
with the modes taken to be orthonormal with
respect to a measure which is just the lattice version of
$\sqrt{\overline g}=\frac{q_i}{3}e^{-2\sigma_i}A_\Delta$. Finally
the overlap of $\sigma_i$ on this set of modes $u_i^{\ell}$ was computed,
using the expression, 
\begin{equation}
\sigma^{\ell} = \int\,d^2x\,\sqrt{\overline g}\, u^{\ell}\,\sigma
\to \sum_i\tilde A_i\,e^{-2 \sigma_i}\,u^{\ell}_i\,\sigma_i\,\ .
\end{equation}  
It is often convenient in numerical calculations to work with dimensionless
quantities $\sigma^{\ell}_{\rm latt}={A_\Delta}^{-\frac{1}{2}}\sigma^{\ell}$ 
and $\lambda_{\ell\,\rm
latt}=A_\Delta\lambda_{\ell}$ corresponding to the operator $A_\Delta L_{ij}$
and lattice measure $\frac{q_i}{3}e^{-2\sigma_i}$.
 
According to the Polyakov-Liouville action $S_{anom}$, the
amplitude of each mode should be distributed as
\begin{equation}
\exp\left\{-{Q^2\over 4\pi}\sum_{\ell}
\lambda_{\ell}\left(\sigma^{\ell}\right)^2\right\} \end{equation}
To check this we rescaled each amplitude $\sigma^{\ell}$ by multiplying
by $(\lambda_{\ell})^{{1\over 2}}$ and calculated the frequency distribution
for this rescaled (dimensionless) amplitude.
All rescaled mode distributions should fit to a Gaussian with constant width determined only by $Q^2$

In addition one can examine the zero mode amplitude $\sigma^0$. This should be
governed by just the linear term in the action $S_{anom}$ (eqn. 5). This distribution is predicted to be (ignoring the fixed area constraint) 
\begin{equation}
P\left(\sigma^0_{\rm latt}\right)\sim  
\exp\left\{-\frac{2Q^2}{\sqrt{N_2}}\sigma^0_{\rm latt}\right\}
\label{zeromod}
\end{equation}
The inclusion of the fixed area constraint (eqn. 14) leads to a more 
complicated distribution for the zero mode. However this constraint has little effect for large amplitudes where we expect a simple exponential behavior.\\

\noindent {\it Numerical Results.}
We show data for both pure 2DQG ($c_m =0$) below the critical 
value of $c_m =1$ and for 2DQG coupled to ten free massless scalar 
fields ($c_m =10$) - well above the critical value. The fits to 
a Gaussian distribution in $\sigma$ are shown in Fig. 1 and Fig. 2
respectively. For $c_m=0$ the fitted width using (rescaled)
mode $\ell =10$ is $w=1.735(10)$ and
corresponds very closely to the value expected from the Liouville action  
$w^L(c_m=0)=\sqrt{\frac{24\pi}{25-c_m}} \simeq 1.7368$. Similarly
the fitted width for (rescaled) mode $\ell =8$ at $c_m=10$ is
$w=2.195(15)$ which lies just 3 standard deviations 
away from its predicted value $w^L(c_m=10)=2.2420$. We attribute the small
discrepancy in the latter to the presence of finite size effects which appear 
somewhat larger for more scalar degrees of freedom.

To examine both the finite size and mode dependence effects more closely
we have plotted in Figs. 3 and 4 the dependence of the rescaled width on 
lattice eigenvalue $\lambda_{\rm latt}$, for volumes $N_2=500$ through
$N_2=1600$. All points correspond to $\chi^2$ per d.o.f. of unity or less. Errors on histograms were obtained using a bootstrap technique. Notice 
first that the (rescaled) widths are relatively insensitive to the eigenvalue 
of the lattice Laplacian and within 10 per cent of so of the Liouville
prediction, which is indicated by the solid horizontal extending away from 
the y-axis. If we focus attention on any curve corresponding to a single 
volume, we notice that for small eigenvalue it climbs steeply to reach a 
broad peak (close to the continuum prediction as indicated by
the solid line). Thereafter it falls very slowly with increasing eigenvalue. Clearly both lattice cut-off effects and finite size effects play a role in determining the precise shape of the curve. This is especially
true deep into the branched polymer phase with $c_m =10$ in Fig. 4,
where the typical geometries are very irregular on the lattice scale $a$. 
We should expect the lattice Laplacian to depart significantly from the continuum there, with the higher modes suffering systematic deviations. 

For the lower eigenvalues the fixed volume constraint which we have 
neglected to this point must also be taken into account. In the continuum partition function this amounts to the insertion of the delta function constraint, $\delta\left(A-\int\sqrt{\overline{g}}e^{\alpha\sigma}\right)$ 
into the functional integral governed by the action (\ref{act}).
In the DT simulation this constraint enters through (\ref{cons}) for the background area. Clearly this constraint couples all the modes nonlinearly
and causes their distribution to depart from a Gaussian, although it is 
natural to expect its effects will be larger on modes with the longest 
wavelengths. In the continuum limit, defined by the process of taking the
average triangle size  $A_\Delta\to 0$ while holding $\bar R$ and the 
continuum eigenvalue fixed, the effect of the constraint becomes unimportant. 
In this limit we should examine the behavior of the eigenvalue curves for small
lattice eigenvalue. The finite size effects enforced by the constraint
(\ref{cons}) prevent us from going literally to zero lattice eigenvalue,
and are responsible for the rapid turn over of the curves at the smallest eigenvalues in Figs. 3 and 4. It is clear that for small enough 
(lattice) eigenvalue the wavelength of a mode saturates at the typical
linear size and hence the {\it rescaled} width goes to zero like the square
root of the lattice eigenvalue. Thus any extrapolation procedure should
utilize the smallest eigenmodes with long wavelengths that
are still significantly smaller than the lattice size. The trend
of the curves in Figs. 3 and 4 with increasing volume indicates
that the modes whose widths lie close to the peak are candidates for 
such `continuum-like' eigenmodes. Indeed the heights of these peaks show 
a convergence to the value expected from the continuum Gaussian theory.

One way to check this conclusion (and the self-consistency of the 
numerical calculations) is to consider the effect of using {\it massive}
rather than massless scalar fields. Fig. 5 shows a plot of the width versus
eigenvalue for a lattice with volume $N_2=1600$ as a function of
scalar field mass $M$. The (lattice) mass parameter was set at $1.0$, $0.1$ 
and finally $0.01$. We see that for lattice eigenvalues $a_\Delta \lambda > 1$
all the curves are statistically similar and compatible with
the data for $M=0.0$ (Fig. 4). In this region of the spectrum the lattice
cannot distinguish massless from massive, and the lattice cut-off effects
dominate. However, at smaller eigenvalues $a_\Delta \lambda < 1$ we
see very different behavior for different masses. The widths for $M=1.0$ start
to differ markedly from $M=0.1$ and $M=0.01$, the latter two
curves themselves splitting apart for small enough eigenvalue 
$a_\Delta \lambda < 0.25$. 
We can understand this behavior by realizing that the scalar
fields will look effectively massless when their correlation length $1/M$ 
is larger than the typical linear extent of the geometries (determined
by the fixed volume). Hence the $M=0.01$ curve is statistically
consistent with the exactly massless data (Fig. 4) over the
entire spectrum. Its peak sits close to the value expected for ten 
massless scalars $c_m=10$. In contrast the data for $M=0.1$ shows a small but
significant deviation from the massless case at the smaller eigenvalues
near the peak, which is consistent with the continuum correlation length
$1/M$ being slightly smaller than the typical linear extent of the geometries. 
In this region of the spectrum the modes have a wavelength
greater than $M^{-1}$ but less than the linear extent of a              
typical geometry in the DT ensemble. The smaller width
of the Gaussian fit for these modes is consistent with
the expectation that the massive modes should begin to
decouple from the trace anomaly in the infrared large volume
limit, yielding an effective value of $c_m$ smaller than
that for $10$ massless scalars.
 For $M=1.0$ the continuum peak is not observed at all for small eigenvalues, the effects of lattice cut-off and
finite volume apparently contaminating the signal even at intermediate wavelengths. Although at this mass
the lattice is a bit too crude to be completely convincing,
the data for $M=1.0$ is more consistent with the width 
expected for pure gravity rather than for gravity coupled to ten massless scalars, suggesting the complete decoupling of these very massive fields from the continuum anomalous effective action (\ref{Sanom}) at this volume. 
For all masses at the largest wavelengths (smallest
eigenvalues) where finite size effects should dominate,
the curves eventually turn over and head towards zero.
 
Thus the massive simulations support our preliminary conclusion
that continuum physics is obtained at small
eigenvalue close to the peak in the width vs. eigenvalue plots,
where both lattice and finite volume effects are minimized.
However it would be fair to say that we do not have a completely
satisfactory quantitative understanding of
these lattice and finite size effects, in the absence of which one may 
question the complete exclusion of the larger eigenvalue region of the plots. 
If the shape of curve is dominated by lattice effects at all
but the longest wavelengths it is reasonable to consider the {\it ratio} of
widths between $c_m=0$ and $c_m=10$ theories over the entire
eigenvalue range, expecting the leading systematic lattice effects
to drop out of this ratio. A plot of this ratio is shown in Fig. 6 which
supports this idea. The fluctuations in the measured ratio vary by only a few percent over all eigenvalues excluding the very lowest. Moreover, the ratio of Gaussian widths is consistent with that expected from the continuum action, $\sqrt{25-10\over 25} \simeq 0.7746$.

Finally the action (\ref{Sanom}) also predicts the asymptotic form of
the zero mode distribution to be a simple exponential, controlled
by the linear term in (\ref{act}), again if the effects of the finite
volume constraint (\ref{cons}) is ignored. From (\ref{cons}) we
expect that the distribution of $\sigma_0$ should become exponentially 
insensitive to the constraint for large $\sigma_0$.
A fit for large values of the zero mode amplitude $\sigma_0$ of 
the logarithm of its distribution for $c_m=0$ is shown in Fig. 7. The data corresponds to a simulation with $N_2=800$ simplices.
The slope of $0.304(10)$ compares well with the expected value from
the continuum, $0.295$. The same mode for $c_m=10$ is also shown
in Fig. 8. Here there are fewer points and their errors are substantially 
larger, consistent with larger finite size effects, but again the
slope of $0.24(2)$ does not differ too badly from
the continuum prediction of $0.177$, lying approximately
three standard deviations away from it.\\

In conclusion, we have shown that is is possible to reconstruct a lattice
conformal mode from the ensemble of DT geometries, even for $c_m>1$.
We have shown that the distribution of this conformal mode is bounded and Gaussian 
with a width corresponding to that expected from the anomalous action 
(\ref{Sanom}), even well into the branched polymer phase. This provides an explicit connection between the ensemble of geometries generated by the DT method and the continuum quantization of 2D gravity. The correspondence between
the DT lattice and continuum approaches demonstrates that the DT measure generates automatically the correct anomaly coefficient $Q^2$ including the Faddeev-Popov ghost contributions in the continuum. This is a non-trivial
result since it implies DTs can be used to study 2D gravity
even in the regime where the KPZ exponents become complex, and the
theory becomes sensitive to its UV cut-off. Finally, the
massive scalar simulations support the proposition that the anomaly
generated action (\ref{Sanom}) is indeed the correct effective Wilsonian
action in the continuum large volume (infrared) limit, where massive states should decouple.\\

\nopagebreak
\samepage
\noindent {\bf Acknowledgments}
The authors would like to thank Tanmoy Bhattacharya for useful discussion.
This work was supported in part by DOE grant  DE-FG02-85ER40237. Simon Catterall would like to thank the T-8 group at Los Alamos National Laboratory where much
of this work was carried out.

\begin{figure}
\epsfxsize=12cm
\epsfysize=9cm
\centerline{\epsfbox{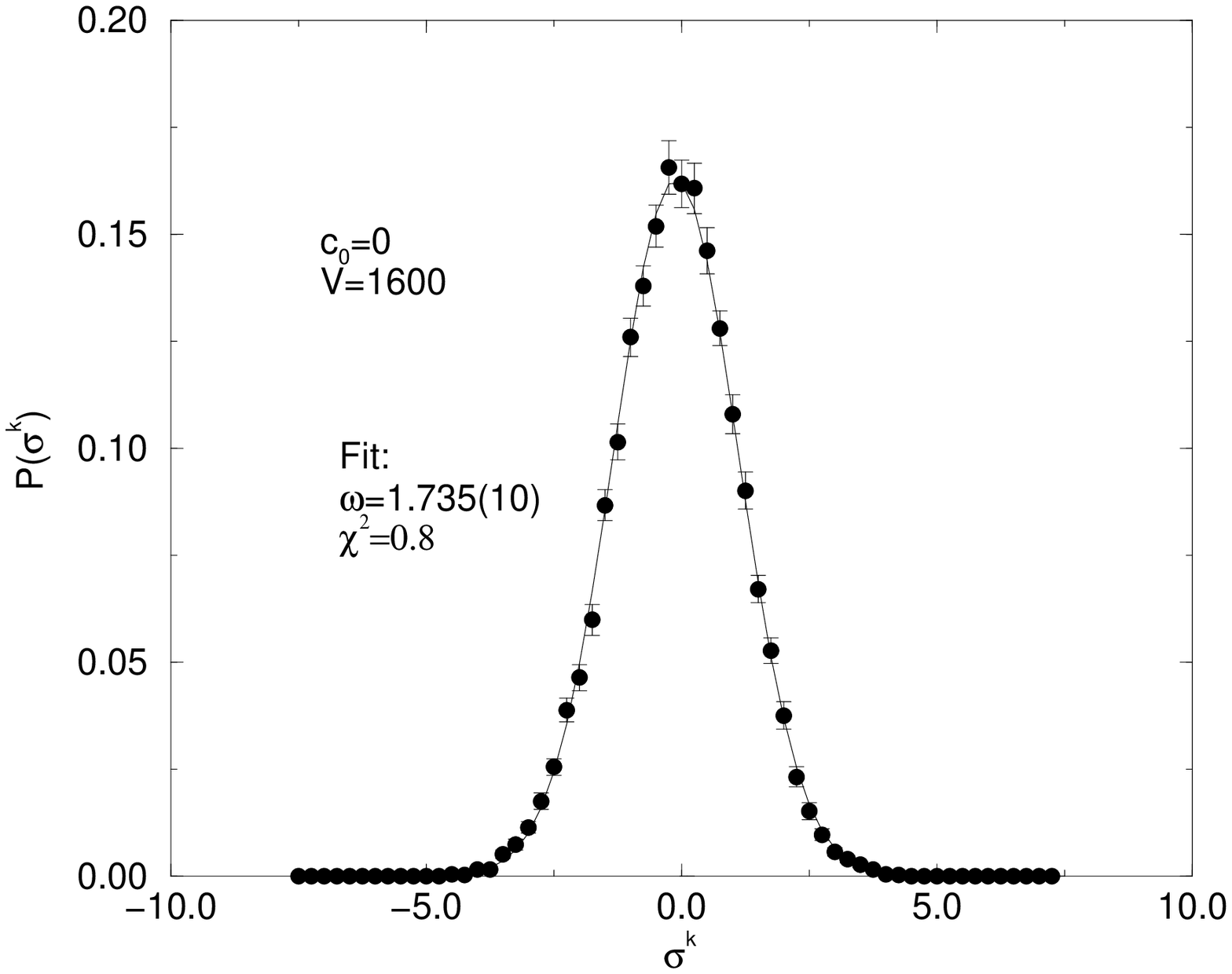}}
\label{Fig. 1}
\caption{Gaussian distribution for (rescaled) amplitude $\sigma^{\ell}$
($\ell=10$) for
pure 2DQG $c_m=0$ in the case of $N_2=1600$ simplices.}
\end{figure}

\begin{figure}
\epsfxsize=12cm
\epsfysize=9cm
\centerline{\epsfbox{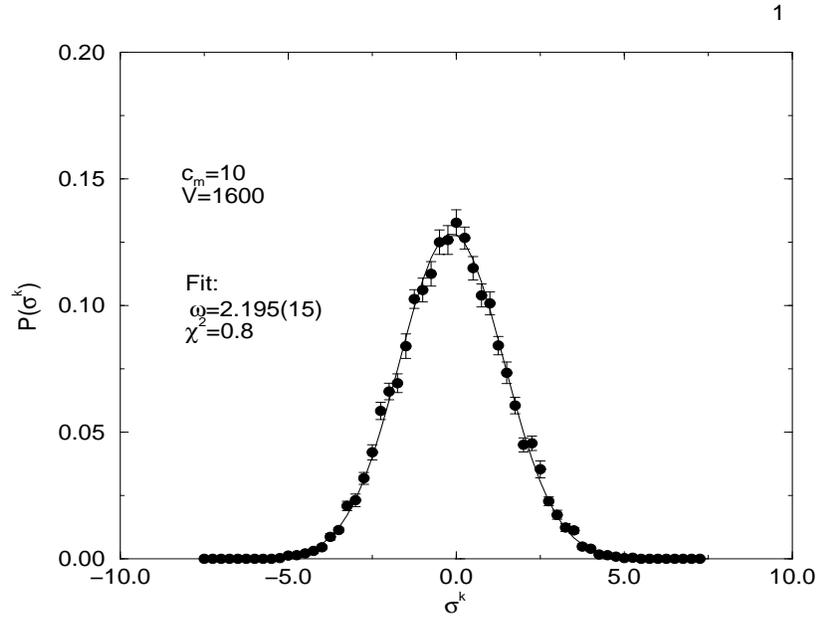}}
\caption{Gaussian distribution for the (rescaled)
amplitude $\sigma^{\ell}$ ($\ell=8$) for 
2DQG coupled to ten free scalar fields ($c_m =10$)
in the case of $N_2 = 1600$ simplices.}
\label{Fig. 2}
\end{figure}

\begin{figure}
\epsfxsize=12cm
\epsfysize=9cm
\centerline{\epsfbox{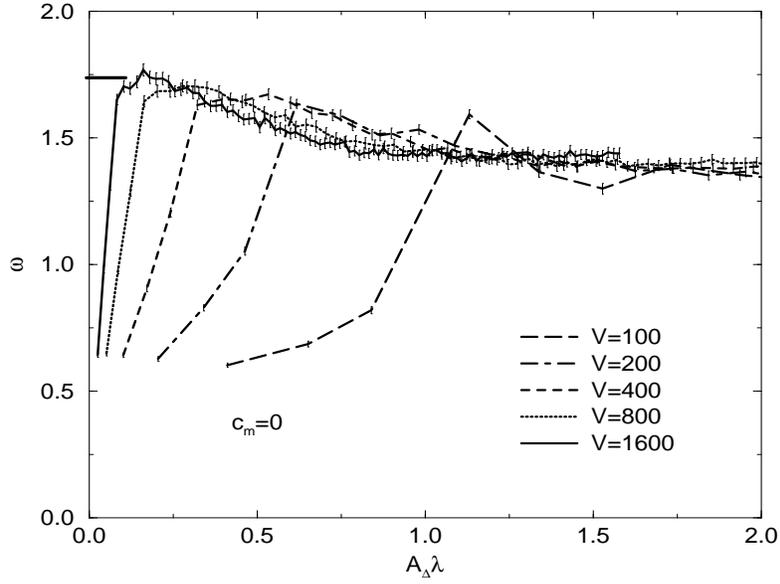}}
\label{Fig. 3}
\caption{The rescaled width of the Gaussian distribution in $\sigma$ vs.
lattice eigenvalue $\lambda_{\rm latt}$ for pure 2DQG ($c_m=0$) in the
case of $N_2=1600$.}
\end{figure}

\begin{figure}
\epsfxsize=12cm
\epsfysize=9cm
\centerline{\epsfbox{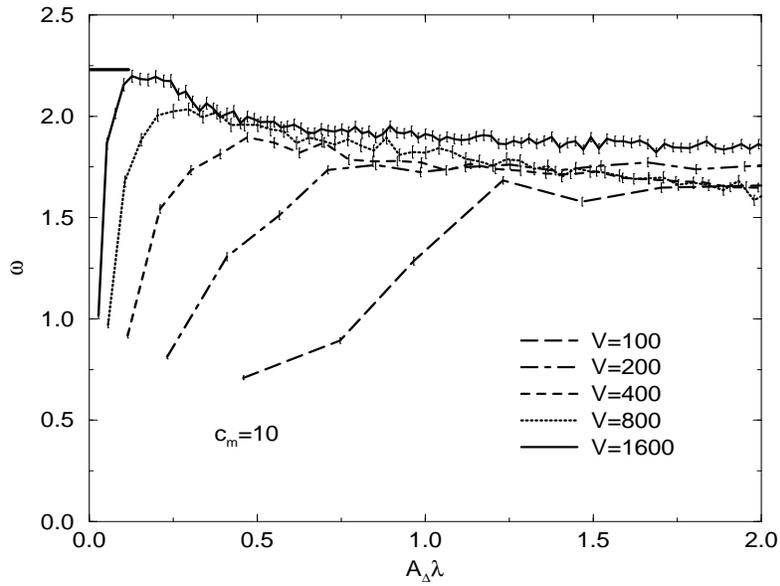}}
\caption{The rescaled width of the Gaussian distribution in $\sigma$ vs.
lattice eigenvalue $\lambda_{\rm latt}$ for 2DQG coupled 
to ten massless scalar fields ($c_m =10$) in the case of 
$N_2 = 1600$ simplices.}
\label{Fig. 4}
\end{figure} 

\begin{figure}
\epsfxsize=12cm
\epsfysize=9cm
\centerline{\epsfbox{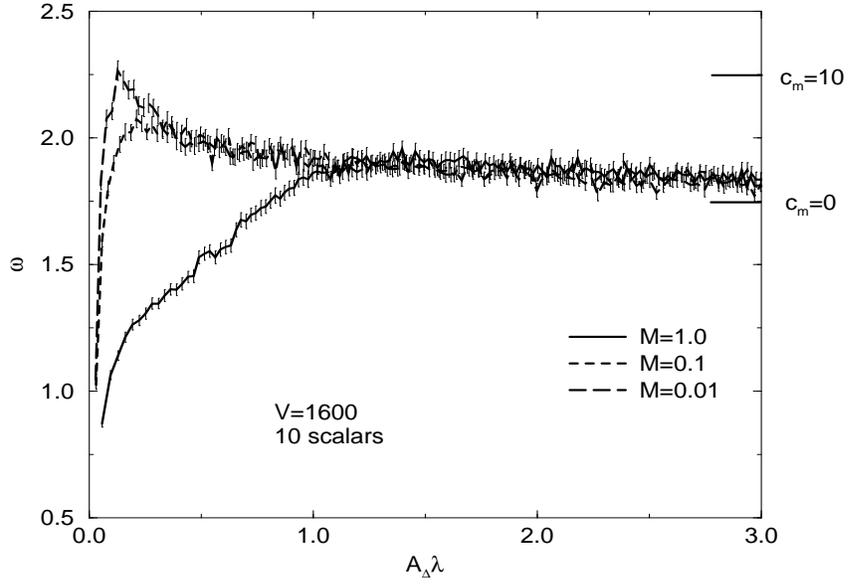}}
\caption{Width vs. eigenvalue for $N_2=1600$ for three different values
of the lattice mass $M=0.01,0.1,1.0$.}
\label{Fig. 5}
\end{figure} 

\begin{figure}
\epsfxsize=12cm
\epsfysize=9cm
\centerline{\epsfbox{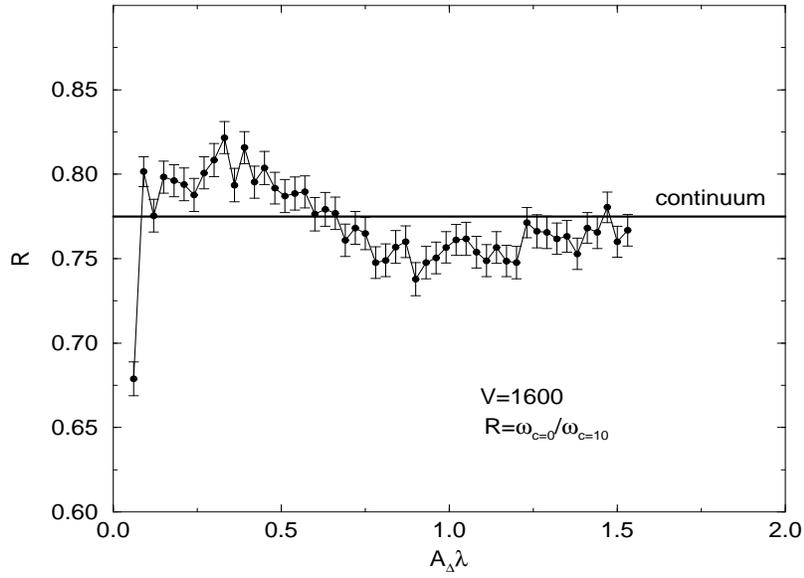}}
\caption{The ratio of Gaussian widths for $c_m =0$ to $c_m=10$
vs. eigenvalue for $V=1600$. The expected continuum ratio of 
$\sqrt{3\over 5}$ is indicated by the solid horizontal line.}
\label{Fig. 6}
\end{figure}

\begin{figure}
\epsfxsize=12cm
\epsfysize=9cm
\centerline{\epsfbox{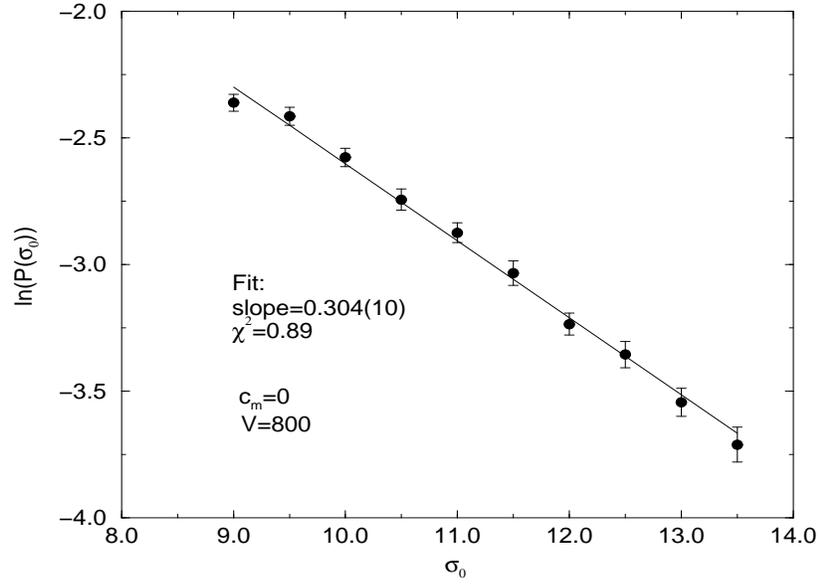}}
\caption{Logarithm of the zero mode distribution in $\sigma^0_{\rm latt}$
vs. amplitude for pure 2DQG ($c_m=0$).}
\label{Fig. 7}
\end{figure}

\begin{figure}
\epsfxsize=12cm
\epsfysize=9cm
\centerline{\epsfbox{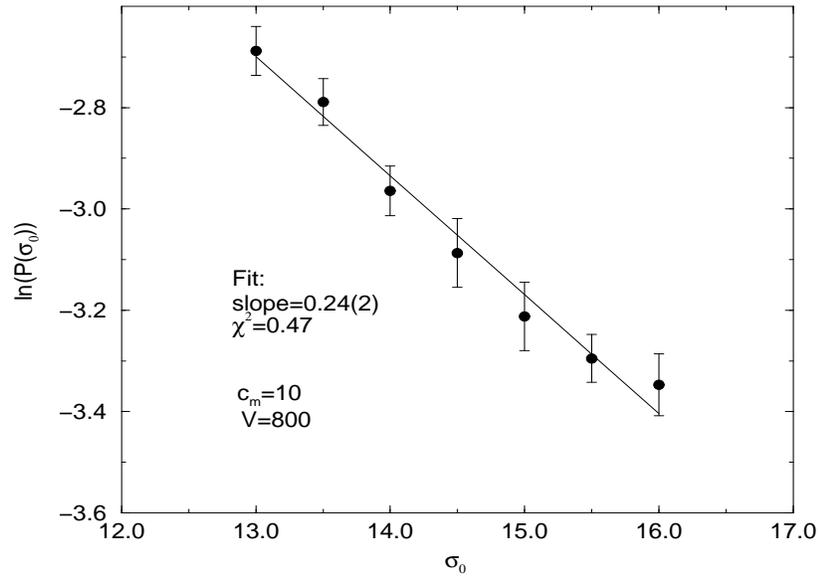}}
\caption{The logarithm of the zero mode distribution in $\sigma^0_{\rm latt}$ vs. amplitude for gravity coupled to $c_m=10$
matter for $N_2 = 800$ simplices. }
\label{Fig. 8}
\end{figure}

\samepage

\end{document}